\documentstyle[12pt]{article}
\begin{document}

\title{Two and three-dimensional spin systems \\
 with gonihedric action}
\date{}

\author{G.K. Bathas$^{(a)}$, E. Floratos$^{(b,c)}$, G.K. Savvidy$^{(b,d)}$,
K.G.Savvidy$^{(e)}$ \\
$^{(a)}$Physics Department, National Technical University of Athens,\\
Zografou Campus, 157-80 Athens, Greece\\
$^{(b)}$Physics Department, University of Crete, 71409 Iraklion, Crete,
Greece\\
$^{(c)}$ $\Delta \eta \mu$\'{o}$\kappa \rho \iota \tau o \varsigma$
National Research Center, P.O. Box $60228$,\\ $15310$ A. Paraskevi, Greece\\
$^{(d)}$Yerevan Physical Institute, 375036 Yerevan, Armenia\\
$^{(e)}$Princeton University, Department of Physics,
P. O. Box 708,\\ Princeton, NJ 08544, U.S.A.}

\maketitle
\begin{abstract}
We perform numerical simulations of the two and three-dimensional spin
systems with competing interaction. They describe the model of random
surfaces with linear-gonihedric action. The degeneracy of the vacuum state
of this spin system is equal to $~~d \cdot 2^{N}~~$ for the lattice of
the size $~N^{d}~$. We observe the second order phase transition
of the three-dimensional system, at temperature $\beta_{c}\simeq 0.439322$
which almost coincides with $\beta_{c}$ of the 2D Ising model. This
confirms the earlier analytical result for the case when self-interaction
coupling constant $k$ is equal to zero. We suggest the full set of order
parameters which characterize the structure of the vacuum states and of
the phase transition.
\end{abstract}

\thispagestyle{empty}
\vfill

\noindent U.Crete TH 18/95

\noindent February 1995

\vspace{.5cm}
\vspace{.3cm}

\newpage

{\bf 1. }

It is well known that the partition function
of three-dimensional Ising
ferromagnet is equivalent to the summation over closed surfaces with an
action which is proportional to the {\it surface area}
\cite{wegner1,waerden,kramers}.
In two and three dimensions the Ising model undergoes a second order phase
transition and at the critical point, in
two-dimensions, corresponds to free fermions
and probably to interacting string theory in three dimensions
\cite{fradkin,itzykson,casher,polyakov,marinari}.

The aim of this article is to investigate the phase structure of the two- and
three-dimensional spin systems introduced in \cite{wegner}.
In three dimensions the
partition function of this spin system  can be
represented as a sum over
closed self-intersecting random surfaces with an action which
is proportional to
the {\it linear size} of the surfaces \cite{ambar,savvidy1,savvidy2}.

We observe the second order phase transition of the system with
linear-gonihedric
action in three dimensions. This numerical result encourages the hope that
the system may describe the propagation of an almost
free fermionic string  \cite{savvidy}.
This hope is based on the fact that the
the $3D$ gonihedric model
can be rewritten as a
model of propagating loops with
the interaction term which is proportional to the
overlapping length, while $3D$ Ising model
can be rewritten  in the same way, but with
the interaction term
which is proportional to overlapping  area, and therefore
in the latter case the interaction is much stronger \cite{savvidy}.

In the next sections we describe the two- and three-dimensional
models with gonihedric
action. We stress  that the vacuum is strongly degenerated and that the
rate of degeneracy depends on the self-intersection coupling
constant $k$. If $k \neq 0$, the degeneracy of the vacuum state is
equal to $~~d \cdot 2^{N}~~$ for the lattice of the size $~N^{d}~$
and is equal to $~2^{dN}~$ when $k=0$. The last case is a sort of
supersymmetric point in the space of gonihedric Hamiltonians
\cite{savvidy3,savvidy}.
The rate of degeneracy
is between Ising model $Z_{2}$ and Wegner
gauge spin systems $~Z_{2}^{N^d}~$.

In two dimensions the system is in completely disordered regime
because of the tunneling phenomenon between the vacuum states
through the instantons. The instantons remove the degeneracy of
the ground state and the symmetry restoreds.

In three
dimensions the vacua are well separated and we have different
symmetry breaking phases. We observed the second
order phase transition at $\beta_{c} \simeq 0.439322$.
We suggest the full set of order parameters
which characterize the structure of the vacuum states and the
phase transition.
At the critical point the correlation length became infinite
and the theory may have a continuum limit.

{\bf 2. }

The Hamiltonian of the system in two dimensions has the form
\cite{wegner,savvidy3}

\begin{equation}
H_{gonimetric}^{2d}=-k \sum_{\vec{r},\vec{\alpha}} \sigma_{\vec{r}}
\sigma_{\vec{r}+\vec{\alpha}}
+ \frac{k}{2} \sum_{\vec{r},\vec{\alpha},\vec{\beta}} \sigma_{\vec{r}}
\sigma_{\vec{r}+\vec{\alpha} +\vec{\beta}}
- \frac{1-k}{2} \sum_{\vec{r},\vec{\alpha},\vec{\beta}} \sigma_{\vec{r}}
\sigma_{\vec{r}+\vec{\alpha}} \sigma_{\vec{r}+\vec{\alpha}+\vec{\beta}}
\sigma_{\vec{r}+\vec{\beta}}
\end{equation}
The low temperature expansion of the partition function

\begin{equation}
Z(\beta)=\sum_{\{\sigma\}}\: exp \: (- \beta\: H^{2d}_{gonimetric}\:)\:
\end{equation}
can be represented as a sum over random walks or
paths with the energy
of the path which is proportional to the number of right corners of the path.
In the self-intersection points  the interaction energy is equal to $4k$ and
therefore the constant $k$
is called an intersection coupling constant \cite{savvidy2,savvidy3}.
The energy of a given path $P$ is equal to
\begin{equation}
E_{path} \equiv k(P)= right(P) + 4 k \cdot inter(P)
\end{equation}
where $right(P)$ is the number of right angles of the path
and $inter(P)$ is the number of
its self-intersection points. $E_{path}$ is a scale invariant quantity: if
we enlarge the contour
let's say $\lambda$ times, then $E_{path}$ does not change.

Comparing this system with $2d$ Ising ferromagnet
\begin{equation}
H_{Ising}^{2d}=- \sum_{\vec{r},\vec{a}} \sigma_{\vec{r}}
\sigma_{\vec{r}+\vec{a}}
\end{equation}
one can see that the energy of paths in the low-temperature
expansions of the partition function is proportional to the length of
the path
\begin{equation}
E_{path} \equiv l(P) =  links(P)
\end{equation}
where $links(P)$ is the number of the links of the closed path.

The intersection coupling constant $k$ defines
the intensity of self-intersections of the paths and at the same time
defines the rate of degeneracy
of the vacuum states. When $k \neq 0$, then
the vacuum is degenerated by
the factor $~d \cdot 2^N~~ $ for the lattice of the size $N^d$, where
$d$ is the dimension of the lattice. The degeneracy increases when $k=0$
and is equal to $2^{dN}$. This means that the point $k=0$ in
the ``space'' of
Hamiltonians corresponds to a system with a higher symmetry.
The rate of degeneracy of the vacuum is between the Ising ferromagnet
and Gauge-spin
models which is  $Z_2$ and $Z_{2}^{N^d}$ correspondingly.

In the case when the intersection coupling constant
is equal to zero ($k=0$) the Hamiltonian
(1) and the partition function reduce to the form
\begin{equation}
H_{gonimetric}^{2d}=- \frac{1}{2} \sum_{\vec{r},\vec{\alpha},\vec{\beta}}
\sigma_{\vec{r}}
\sigma_{\vec{r}+\vec{\alpha}} \sigma_{\vec{r}+\vec{\alpha}+\vec{\beta}}
\sigma_{\vec{r}+\vec{\beta}}
\end{equation}

\begin{equation}
Z(\beta)=\sum_{\{\sigma\}}\: exp \: (- \beta\: H^{2d}_{gonimetric}\:)\: =
\sum_{\{ P \} } \: \: e^{- \beta\: k(P)},
\end{equation}
where $k(P)$ is  the total curvature of the path (3).
In this particular case $k=0$ the model can be solved exactly \cite{savvidy}.
Indeed the high temperature expansion of the partition function is:
\begin{equation}
Z(\beta)=\sum_{\{\sigma\}}\: exp \: (- \beta\: H^{2d}_{gonimetric}\: ) \: =
(\: 2 \: ch \beta )^{N^{2}} (1+ {\cal O}(N)+ \ldots )
\end{equation}
and the free energy is equal to
\begin{equation}
- \beta f=ln(2ch\beta)
\end{equation}
The system is in a completely disordered regime like the one-dimensional Ising
model with energy density and the specific heat equal to:
\begin{equation}
u=\frac{<k(P)>}{N^2}=\frac{1-th\beta}{2},~~~~~C=\beta^2 (1-th^2 \beta) .
\end{equation}
The reason why the system is in the disordered regime is connected with
the fact that
the potential barrier between these vacua is finite like in the
$1D$ Ising model and the tunneling phenomenon
through the instantons removes the degeneracy of
the ground state and the symmetry restores.

The question to which we would like to address
here is the nature of the phase
transition in the case $k \neq 0$. The model with $k \neq 0$
can be formulated as
an eight vertex model with the weights \cite{savvidy3}
\begin{equation}
\omega_1 =1,\; \omega_2 = \omega^{4k}, \omega_3 = \omega_4 =1,
\omega_5 = \omega_6 = \omega_7 = \omega_8 = \omega
\end{equation}
where $\omega = exp(- \beta)$, and the partition function can be
rewritten as a
fermionic partition function with quadratic and quartic field operators.
Perturbation expansion around quadratic operators ($k=0$) shows
that the system is in the disordered
regime for $small$ values of intersection coupling constant $k$.

To confirm this universal behavior of the system for
different values of the self-intersection coupling constant
$k$ one can use the Monte-Carlo simulation of the
spin system (1). Let us consider the case $k=1$ \cite{wegner}.

\begin{equation}
H^{2d}_{gonimetric}=- \sum_{\vec{r},\vec{\alpha}}
\sigma_{\vec{r}}
\sigma_{\vec{r}+\vec{\alpha}}
+ \frac{1}{2} \sum_{\vec{r},\vec{\alpha},\vec{\beta}}
\sigma_{\vec{r}}
\sigma_{\vec{r}+\vec{\alpha}+\vec{\beta}}
\end{equation}
We performed a simulation of the system on a $100^2$ lattice
with periodic boundary conditions.
We measured the magnetization ${\cal M}= < \sigma >_{\beta}$, the spin-spin
correlation function and the average energy density.
We saw that in two dimensions
there is no phase transition
for any positive temperature in accordance with the arguments above.

In order to measure magnetization we start from a high value of $\beta$
with a completely magnetized state ${\cal M}=1$. While we heat up the system,
magnetization drops and the remaining fluctuations are around ${\cal M}=0$.
As we cool down again, we see that the system does become
magnetized spontaneously.
This is the first similarity encountered between the $2D$ gonimetric model and
$1D$ Ising. The second similarity
comes from the measurements of the spin-spin correlation function
${\bf G} ( \vec{r} )
= < {\bf \sigma} (0), {\bf \sigma} ( \vec{r} ) >$ which was
seen to drop very
fast with distance in the large temperature interval.
Our measurements of the critical exponent $\nu$ for the
correlation length $\xi$ at $\beta_c = \infty $
indicate that $\nu$ is close to one.
The third similarity observed is
that typical configurations for high
$\beta$ have a simple structure: the islands formed by spins are
rectangular blocks, the system becomes more and more ordered
as $\beta$ becomes higher. At high temperature
the islands have curly boundaries, but still the
intersections are very disfavoured.
This is a consequence of the high energy cost associated with them.

Again one can explain this disordered phase by analyzing the vacuum structure.
As we already mentioned in the case of Hamiltonian $(1)$
with $k \neq 0$ the vacuum is $2 \cdot 2^N$ times
degenerated, because flat layers of spins with opposite directions have zero
interface energy.
The vacuum state of the 2D Ising model is twice degenerated and the
potential barrier between those vacua  is infinite, while in the
gonimetric model $(1)$ the barriers between the vacuum states are finite
and are equal to four units of
energy. This is the reason why the system is always in a disordered regime.
The kinklike excitations which connect those vacua
are shown on Fig. 1.

In summary for the two-dimensional system with gonimetric
action we have the
following properties:
(i) The system has strong degeneracy of the vacuum states, similar to
gauge or spin glass systems, which
are separated by finite potential barriers.
There is a tunneling
phenomenon between those vacua through instantons.
(ii) Consequently, the instantons remove the degeneracy of
the ground state and the symmetry restoreds.
The system is in the disordered regime for
 all non-zero temperatures and for large
values of the intersection coupling constant $k$.
(iii) The typical configurations of interface
are such that the paths of interface are self-avoiding.

{\bf 3. }

In three dimensions the corresponding Hamiltonian is \cite{wegner,savvidy3}
\begin{equation}
H_{gonihedric}^{3d}=- 2k \sum_{\vec{r},\vec{\alpha}} \sigma_{\vec{r}}
\sigma_{\vec{r}+\vec{\alpha}}
+ \frac{k}{2} \sum_{\vec{r},\vec{\alpha},\vec{\beta}} \sigma_{\vec{r}}
\sigma_{\vec{r}+\vec{\alpha} +\vec{\beta}}
-  \frac{1-k}{2} \sum_{\vec{r},\vec{\alpha},\vec{\beta}} \sigma_{\vec{r}}
\sigma_{\vec{r}+\vec{\alpha}} \sigma_{\vec{r}+\vec{\alpha}+\vec{\beta}}
\sigma_{\vec{r}+\vec{\beta}},
\end{equation}
and the low temperature expansion of the partition function
\begin{equation}
Z(\beta)=\sum_{ \{ \sigma \}} exp( - \beta H_{gonihedric}^{3d} )
\end{equation}
can be expressed as a sum over surfaces of interface with
the amplitude which is
proportional to linear size of the surface \cite{ambar,savvidy1,savvidy2}.
\begin{equation}
Z(\beta)=\sum_{ \{ M \} } e^{ - \beta {\cal A}(M)} \quad ,
\quad {\cal A}(M)= \sum_{<i,j>}
\lambda_{ij} \cdot \vert \pi - \alpha_{i,j} \vert
\end{equation}
where $\lambda_{ij}$ is the length of the surface edges
(on the lattice $\lambda$ is
equal to lattice size $a$), and $\alpha_{ij}$ is the dihedral angle
equal to $0~,\pi/2 $ or $\pi $ .

The constant
$k$ again plays the role of intersection coupling constant
and defines the rate of degeneracy of the vacuum.
If we take periodic boundary conditions and $k=0$,
then the vacuum is degenerated by the factor
$3\cdot 2^N$~~ and consists of spin layers  of different
width  in the x, y, and z
direction.
In the same way as in two dimensions the system simplifies
in the symmetric point where the
self-intersection coupling constant is zero $k=0$ \cite{savvidy3,savvidy}
\begin{equation}
H_{gonihedric}^{3d}=- \frac{1}{2} \sum_{\vec{r},\vec{\alpha}, \vec{beta}}
\sigma_{\vec{r}}
\sigma_{\vec{r}+\vec{\alpha}}
\sigma_{\vec{r}+\vec{\alpha}+\vec{\beta}}
\sigma_{\vec{r}+\vec{\beta}}
\end{equation}
The vacuum is strongly degenerate $~2^N \cdot 2^N \cdot 2^N~~ $ times ,
because now all
flat layers in all directions form different vacuum states.
The system (17) is highly symmetric and even allows to construct the
dual Hamiltonian \cite{savvidy} (the point $k=0$ in the
space of Hamiltonians).

Little is known about phase transition point of the three-dimensional
system (14) even in supersymmetric point $k=0$ (17) .
Curvature representation of the linear
action (16) \cite{schneider} allows to find
equivalent representations of the partition function for the
system (17) in terms
of propagation of the
polygon loops or strings $P$ in a given direction with the
transition amplitude \cite{savvidy}
\begin{equation}
exp \; ( - k(P) -2\cdot l(P))
\end{equation}
where $l(P)$ is the length and $k(P)$ is the absolute curvature of $P$.
The interaction is proportional to the overlapping length of the loops
\begin{equation}
A_{int} = l( P_1 \cap P_2 ).
\end{equation}
In the first approximation ignoring the interaction term one can
solve the model
and see that the system undergoes a second order
phase transition at $\beta_c$ which is  similar to
$2D$ Ising ferromagnet and that it describes the propagation
of an almost free string of $2D$ Ising fermions \cite{savvidy}.

One can understand the nature of this second order phase
transition analyzing the structure of the vacuum. Indeed
the potential barrier between those vacua is infinite like in
2D Ising model and is of order $N$ for the lattice of the size
$N^{3}$. Therefore one can have  well separated phases of
the system at low temperature, but rigorous prove of there
existence is still missing. The main problem is to
count properly the entropy of the given configurations.

This symmetry breaking phenomenon cannot be uniquely
characterized by an order
parameter like magnetization, because the
number of vacuum states with zero magnetization is equal to
\begin{equation}
\frac{N!}{(( \frac{N}{2})!)^2} \simeq \frac{2^N}{ \sqrt N }
\end{equation}
and the probability to have zero magnetization dominates.
The degenaracy is less for the vacuum states with nonzero
magnetization.
The magnetization does not
uniquely characterize the system and only partly play the
role of an order
parameter of the system.
One can introduce the full set of order parameters to characterize
this phases. Let as denote the vacuum spin comfigurations
of the system by
$\sigma^{\mu}_{\vec r}(vac)$ where $\mu =1,2,...,2^{3N}$, then
the generalized magnetization $M^{\mu}$ is equal to
\begin{equation}
M^{\mu} = <\sum_{\vec r}\sigma^{\mu}_{\vec r}(vac)
\cdot \sigma_{\vec r}>
\end{equation}
and uniquely characterize the vacuum statets and the phase transition.
The internal energy, which is
equal to the number of corners can also be used as an
order parameter.

It is interesting therefore to have Monte-Carlo simulations
of the system in three
dimensions and to study the nature and the dependence of
the phase transition as a function of the
self-intersection coupling constant $k$ in $(14)$.
We will consider here the case $k=1$ \cite{wegner}, when
\begin{equation}
H_{gonihedric}^{3d}=- 2 \sum_{\vec{r},\vec{\alpha}} \sigma_{\vec{r}}
\sigma_{\vec{r}+\vec{\alpha}}
+ \frac{1}{2} \sum_{\vec{r},\vec{\alpha},\vec{\beta}}
\sigma_{\vec{r}} \sigma_{\vec{r}+\vec{\alpha}+\vec{\beta}}
\end{equation}
and the spin system includes only the competing
ferromagnetic and antiferromagnetic interaction.

We consider the
$3D$  lattice of size $N^3 = 24^3$ and $32^3$. Measuring the energy density
\begin{equation}
u(\beta)=\frac{<{\cal A}(M)>}{N^3} =
\frac{\partial}{\partial \beta} (\beta f(\beta))
\end{equation}
we observe the sharp behavior at temperature $\beta_c \simeq 0.439322$,
see Fig. 2.
The two point correlation function drastically
increase  at critical temperature, see Tab.1.
As we approach the transition point from higher temperatures
the correlation function blows-up uniformly in all three directions.
Afther critical point, at low temperatures, the correlation function
is unisotropic and we observe symmetry breaking phenomena.
All energy histograms obtained near $\beta_{c}$ have only one pick,
see Fig. 3.
The typical spin configurations, while we set near critical point ,
describe large surfaces without self-intersections.

We came to the conclusion that the system in three dimensions undergoes
a second
order phase transition and therefore may have well defined
continuum limit.
For a rigorous proof of this picture one should have a more detailed
study of the phenomena near the critical point.
We still dont know the critical indices of the model to
distinguish them from the ones in 3D Ising model.

\vspace{.5cm}
{\bf Acknowledgements.} We have benefited from
fruitful discussions with Franz Wegner.
We are thankful to G.Athanasiou for the discussion of
the spin glass systems.
G. Bathas would like to thank the Physics Department
of the University of Crete,
where this work was performed, for hospitality.

\vfill
\newpage
\vspace{5cm}
{\bf Figure caption}

{\bf Fig. 1.} (a) one of the vacuum states,
(b) excitation - the kink of energy four,
(c) new vacuum after transition .

{\bf Fig. 2.} Total number of corners as a function of
$\beta$ in $32^{3}$~~~ 3d lattice, at $k=1$.

{\bf Tab. 1} Corralation function at $\beta = 0.439320$.

{\bf Fig. 3.} Energy histogram at $\beta = 0.439322$

\end{document}